\documentclass[reprint,aps,longbibliography,nofootinbib,floatfix,nobalancelastpage]{revtex4-2}
\usepackage{amsmath}
\usepackage{amsfonts}
\usepackage{amssymb}
\usepackage{mathtools}
\usepackage{graphicx}
\usepackage[dvipsnames]{xcolor}
\usepackage[colorlinks=True,citecolor=gray,linkcolor=myRed,urlcolor=gray]{hyperref}
\usepackage{hypcap}
\usepackage{yfonts}
\usepackage{bm}
\usepackage{dsfont}
\usepackage{braket}

\newcommand\eq[1]{\begin{align}#1\end{align}}

\newcommand\nh{N_\mathcal{H}}

\newcommand{\pu}{\mathcal{P}^{AB}}

\newcommand{\z}{\sigma^z}
\newcommand{\al}{\alpha}
\newcommand{\be}{\beta}
\newcommand{\ga}{\gamma}
\newcommand{\la}{\lambda}

\newcommand{\pos}{\tilde{\cal P}_\square(l,\omega)}
\newcommand{\posrArB}{\tilde{\mathcal{P}}_{\square_{r_Xr_Y}}(\omega)}

\newcommand{\tabcd}{\theta_{\alpha\beta\gamma\lambda}}

\newcommand{\Gabcd}{\mathcal{V}^{(l)}_{\al\be\ga\la}}
\newcommand\Vabcdl[1]{V^{(#1)}_{\alpha\beta\gamma\lambda}}

\definecolor{myBlue}{RGB}{31,119,180}
\definecolor{myOrange}{RGB}{255,127,14}
\definecolor{myGreen}{RGB}{44,160,44}
\definecolor{myRed}{RGB}{214,39,40}
\definecolor{myPurple}{RGB}{148,103,189}

\makeatletter
\def\p@figure{\color{myBlue}}
\def\p@equation{\color{myRed}}
\makeatother

\begin{document}

\title{Logarithmic entanglement lightcone from eigenstate correlations in the many-body localised phase}

\author{Ratul Thakur}
\email{ratul.thakur@icts.res.in}
\affiliation{International Centre for Theoretical Sciences, Tata Institute of Fundamental Research, Bengaluru 560089, India}

\author{Bikram Pain}
\email{bikram.pain@icts.res.in}
\affiliation{International Centre for Theoretical Sciences, Tata Institute of Fundamental Research, Bengaluru 560089, India}

\author{Sthitadhi Roy}
\email{sthitadhi.roy@icts.res.in}
\affiliation{International Centre for Theoretical Sciences, Tata Institute of Fundamental Research, Bengaluru 560089, India}

\begin{abstract}
We investigate the operator entanglement of the time-evolution operator through the framework of eigenstate correlations. Focusing on strongly disordered quantum many-body systems in the many-body localised (MBL) regime, we analyse the operator entanglement across various spatiotemporal cuts, revealing the logarithmic lightcone of entanglement spreading. We demonstrate that this logarithmic lightcone arises directly from a hierarchy of energyscales and lengthscales encoded in eigenstate correlations. By characterising the statistics of these hierarchical scales, we develop a microscopic theory for the spatiotemporal structure of entanglement spreading in MBL systems—without invoking phenomenological constructs such as $\ell$-bits. This approach reveals the fundamental connection between eigenstate correlations and the emergent entanglement structure in MBL systems.
\end{abstract}

\maketitle

\section{Introduction}

The spatiotemporal profile of quantum correlations and information in out-of-equilibrium quantum many-body systems has been a topic of central interest of late in condensed matter and statistical physics~\cite{dalessio2016from,nandkishore2015many,abanin2019colloquium,fisher2023random}. 
This interest has been fuelled in large part due to the increasing cross-pollination of ideas across the fields of quantum information science and quantum many-body physics. This is also due to the capabilities of modern experimental platforms to study such information-theoretic measures in quantum systems over hitherto unexplored timescales~\cite{blatt2012quantum,islam2015measuring,kaufman2016quantum,gross2017quantum,brydges2019probing,arute2019quantum,altman2021quantum,Lukin2019entanglementmbl,fauseweh2024quantum}.
While quantum information theoretic ideas have helped discover new kinds of phases as well as their classifications in many-body systems, fundamental ideas from quantum many-body physics have aided the development of novel quantum computing platforms and quantum algorithms therein~\cite{preskill2018quantum,ippoliti2021many}.
A remarkable outcome of this cross-pollination has been the emergence of quantum entanglement as one of the most important players across the two fields~\cite{eisert2006entanglement,laflorencie2016quantum}. 
In the context of quantum many-body systems, entanglement has now become a new paradigm for classifying states and phases of quantum matter, both in equilibrium~\cite{eisert2010area,wen2017zoo,zeng2019quantum} and out of equilibrium, based on how the states encode quantum information~\cite{bauer2013area,nandkishore2015many,abanin2019colloquium,laflorencie2022entanglement,potter2022entanglement,fisher2023random}.

In isolated quantum systems undergoing time-evolution, the dynamics of entanglement goes hand-in-hand with the fundamental notion of if the system thermalises or not under the dynamics, and how does quantum information spread therein~\cite{deutsch1991quantum,srednicki1994chaos,tasaki1998from,rigol2008thermalisation,dalessio2016from,nandkishore2015many,popescu2006entanglement,linden2009quantum}. 
The most ubiquitously studied measure of entanglement in such settings is the bipartite entanglement entropy between two parts of the system.
In ergodic systems, thermalisation is typically accompanied by a ballistic growth of the bipartite entanglement entropy~\cite{kim2013ballistic}. 
On the other hand, in systems with robustly broken ergodicity, such as in localised systems, the entanglement growth is either completely suppressed or ultraslow~\cite{znidaric2008many,bardarson2012unbounded,serbyn2013universal,nanduri2014entanglement,kim2014localintegralsmotionlogarithmic,Lukin2019entanglementmbl,huang2021extensive,toniolo2024dynamicsmanybodylocalizedsystems}.
While entanglement entropy contains signatures of the temporal behaviour of information spreading, it does not betray anything about the spatial structure of the same.

On this front, the out-of-time ordered correlators (OTOCs) have surfaced as useful probes for the spatiotemporal pattern of information spreading. They exhibit a linear lightcone structure in ergodic systems~\cite{aleiner2016microscopic,luitz2017information,bohrdt2017scrambling,chan2019eigenstate,foini2019eigenstate,keyserlingk2018operator,nahum2018operator,khemani2018velocity} whereas in localised systems, the lightcone spreading is either completely arrested or anomalously slow~\cite{chen2016universal,chen2017out,huang2017out,deng2017logarithmic}. 
While undoubtedly insightful, and also practically useful for ergodic systems, OTOCs raise some questions when applied to strongly disordered systems where there is a sense of a `preferred' basis. This is due to the fact that degrees of freedom to which the disorder couples may behave differently to those to which the disorder does not couple. This leads to a strong dependence on the choice of operators in the results for the OTOCs ~\cite{chen2017out,huang2017out}. 

It is therefore of natural interest to try to understand the spatiotemporal profile of information spreading without alluding to the dynamics of specific operators and just from inherent properties of the system, such as the spectral and eigenstate correlations of the Hamiltonian (or the generator of time-translation in general).
The operator entanglement entropy (opEE)~\cite{prosen2007operator,pizorn2009operator,zhou2017operator,jonay2018coarsegraineddynamicsoperatorstate} of the time-evolution operator befits this purpose as (i) it can be shown to contain signatures of both, the spatial as well as the temporal profile of information scrambling and (ii) it can be represented straightforwardly in terms of specific correlations between (four) different eigenstates and the corresponding eigenvalues~\cite{hahn2023statistical,pain2024entanglement}.

In this work, we study the opEE of the time-evolution operator through the lens of eigenstate correlations for strongly disordered quantum systems in the many-body localised (MBL) regime. The motivation for focussing on the MBL regime is multifold. First, it constitutes a rather unusual setting where the system does not thermalise and there is no transport of conserved quantities~\cite{basko2006metal,gornyi2005interacting,oganesyan2007localisation,pal2010many,nandkishore2015many,abanin2017recent,abanin2019colloquium,alet2018many,sierant2024manybody}, and yet information spreads albeit very slowly in spacetime. Second, the ultraslow entanglement growth in MBL systems~\cite{bardarson2012unbounded,serbyn2013universal,nanduri2014entanglement,kim2014localintegralsmotionlogarithmic} was understood for long using a phenomenological $\ell$-bit picture~\cite{serbyn2013local,huse2014phenomenology} with no complete microscopic theory. This was until recently, when a microscopic picture for the phenomenon was developed based on eigenstate correlations and hierarchy of energyscales and lengthscales therein~\cite{pain2024entanglement}. However, Ref.~\cite{pain2024entanglement} addressed only the temporal growth of entanglement.
In this work we extend the ideas to spatiotemporal profiles through the opEE. We show that the same hierarchy of energy- and lengthscales can be used to develop a microscopic theory for the spatiotemporal profile of the opEE from which the logarithmic lightcone of quantum information falls out automatically. This constitutes the central result of this work.

The rest of the paper is organised as follows. In Sec.~\ref{sec:opEE-eig-corr} we define the opEE of the time-evolution operator with the general setup and definitions introduced in Sec.~\ref{sec:opEE}, and the relation between the opEE and eigenstate correlations discussed in Sec.~\ref{sec:eigcorr}. Numerical results for the opEE and its scaling behaviour for a disordered, interacting Floquet spin-1/2 chain are presented in Sec.~\ref{sec: Num-Floquet-Ising}. In Sec.~\ref{sec:log-lightcone-eigcorr} we show how the logarithmic entanglement lightcone emerges out of a hierarchy of energyscales in the eigenstate correlations. Specifically, in Sec.~\ref{sec:special-quartets} we identify special sets of eigenstates spectral correlations within which dominate the dynamics of the opEE, the statistics and scaling of these spectral correlations are discussed in Sec.~\ref{sec:spect-corr}, and in Sec.~\ref{sec:log-lc-deriv} we present the derivation of the logarithmic lightcone. We close with concluding remarks in Sec.~\ref{sec:conclusions}.

\section{Operator Entanglement and Eigenstate correlations}\label{sec:opEE-eig-corr}

\begin{figure}
\includegraphics[width=\linewidth]{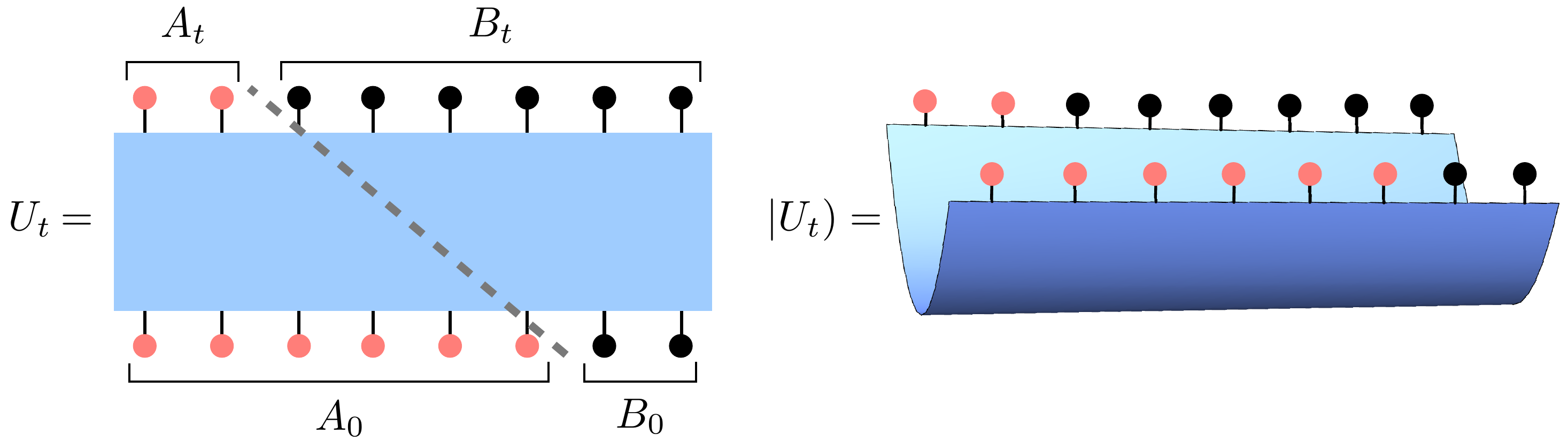}
\caption{Operator entanglement as the entanglement in a state defined on twice as many sites. The left panel shows the unitary operator $U_t$; the legs at the bottom denote the sites which make up the `input' states whereas the those at the top denote the sites which form the `output' state. The bipartitions at $t=0$ and at $t=t$ are also labelled. Right: The operator is  folded such that it can be interpreted as a state (all legs in the same direction) on twice as many sites. The spatiotemporal bipartition on the left can now be identified as a bipartition between sites coloured orange and those coloured black.}
\label{fig:Ut-schem}
\end{figure}

\subsection{opEE of the time-evolution operator \label{sec:opEE}}

We start with describing the opEE of the time-evolution operator and laying out the basic definitions.
Let us denote the unitary operator which effects time-evolution until time $t$ as $U_t$. The operator can be formally written as 
\eq{
U_t = \sum_{i_0,i_t}U_{i_t i_0} (t)\ket{i_t}\bra{i_0}\,,
\label{eq:Ut-gen}
}
where $\{\ket{i_0}\}$ denotes a set of basis states at the initial time and similarly for $\{\ket{i_t}\}$. 
For a system with $L$ sites, the operator can be equivalently viewed as a state (after suitable normalisation) of $2L$ qubits, with $L$ of them at $t=0$ and $L$ of them at $t=t$, for which we use the notation
\eq{
|U_t) = \frac{1}{\sqrt{\nh}}\sum_{i_0,i_t}U_{i_t i_0} (t)\ket{i_t}\otimes\ket{i_0}\,,
\label{eq:U-state}
}
where $\nh$ is the dimension of the original Hilbert space of the $L$-site system, denoted by ${\cal H}$, and $|U_t)$ is a state in the doubled Hilbert space ${\cal H}\otimes {\cal H}$ with dimension $\nh^2$. The $1/\sqrt{\nh}$ factor in Eq.~\ref{eq:U-state} ensures normalisation~\footnote{The inner product of two operators ${\cal X}$ and ${\cal Y}$ in this doubled Hilbert space is defined as $({\cal{X}|\cal{Y}}) = \frac{1}{\nh}{\rm Tr}[{\cal X}^\dagger {\cal Y}]$.
} 
$(U_t|U_t) = \nh^{-1}{\rm Tr}[U^\dagger_tU_t] = 1$.

The mapping of the operator to a state in the doubled Hilbert space then allows for an interpretation and the computation of the opEE in the same way as for entanglement of states. Consider a bipartition of the $2L$ qubits as shown in Fig.~\ref{fig:Ut-schem} where the subsystem ${A=A_0\cup A_t}$ is comprised of the subsystem $A_0$ at $t=0$ and the subsystem $A_t$ at $t=t$, and similarly the subsystem ${B=B_0\cup B_t}$.
The opEE between $A$ and $B$, in particular, the second R\'enyi entropy of operator entanglement is defined as
\eq{
S_2^{AB}(t) = -\ln {\rm Tr}_A\left[\rho^2_A({U_t})\right]\,,
\label{eq:S2-opEE}
}
where $\rho_A(U_t) = {\rm Tr}_B[|U_t)(U_t|]$ is the reduced density matrix of $A$. We will in fact, focus on the operator purity 
\eq{
\pu(t) = \exp[-S_2^{AB}(t)] = {\rm Tr}_A\left[\rho^2_A({U_t})\right]\,.
\label{eq:pu-opEE}
}

While the above framework is rather general, we now focus on the case of one-dimensional systems and a class of bipartitions which probe the spatiotemporal structure of information scrambling.
In particular, consider a bipartition, parametrised by $l$, where $A_t$ contains the $L/2-l$ leftmost sites and $A_0$ consists of the $L/2+l$ leftmost sites, see Fig.~\ref{fig:Ut-schem}.
Note that $l=0$ corresponds to a more ubiquitously studied half-chain operator entanglement~\cite{zhou2017operator} whereas  varying $l$ at a fixed $t$ probes the spatial pattern of information spreading. We will denote the operator purity of $U_t$ for a cut labelled by $l$ as ${\cal P}(l,t)$ and the entire spatiotemporal pattern of information scrambling is encoded in the $l,t$ dependence of ${\cal P}(l,t)$. To see this we consider some some limiting cases on general grounds.

\begin{figure}
    \centering
    \includegraphics[width=\linewidth]{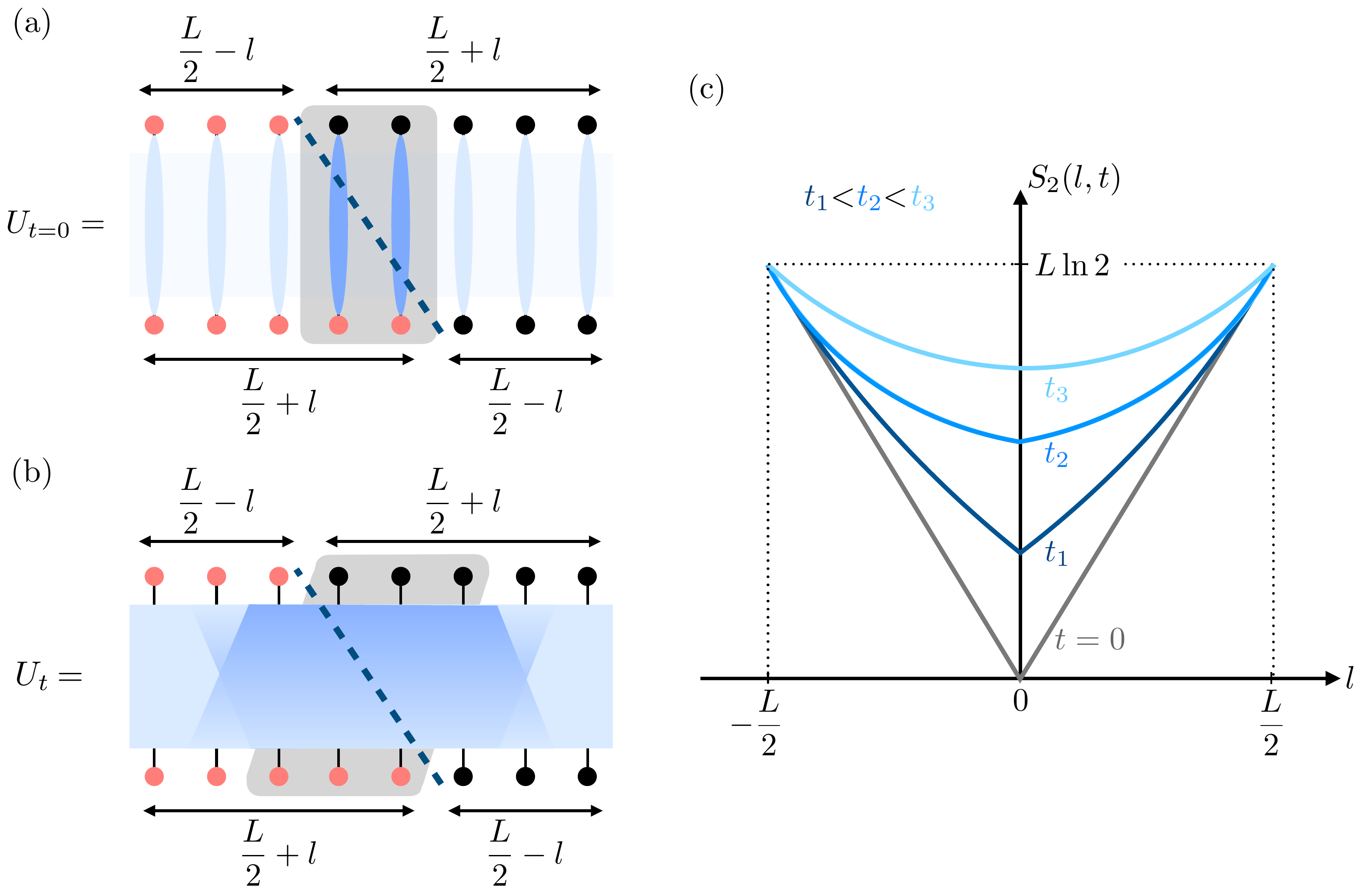}
    \caption{Schematic representation the opEE of $U_t$, $S_2(l,t)$. Left: The dashed line shows the spatiotemporal cut labelled by $l$, with the number of sites in each subsystem marked explicitly. At $t=0$ (a), the operator $U_{t=0}={\mathbb{I}}$ represented as a state in the doubled Hilbert space is a direct product state of Bell pairs between the spins at the same site. As such, the opEE is simply determined by the number of Bell pairs that the spatiotemporal cut goes through -- the grey shaded box denotes those Bell pairs. At $t>0$ (b) spins at any site at $t=0$ get entangled with the spins at other sites at $t=t$ within a lightcone such that the opEE grows -- the grey shaded box denotes the sites which contribute to the opEE for the given cut. (c) Schematic picture for $S_2(l,t)$ as a function of $l$ for different $t$.}
    \label{fig:schematic-operator-entanglement}
\end{figure}

Initially, $U_{t=0}=\mathbb{I}$, whose state representation in the doubled Hilbert space is
\eq{
|U_{t=0}) = \frac{1}{\sqrt{2^L}}\bigotimes_{x=1}^L[\ket{\uparrow_{t=0}}_x\ket{\uparrow_0}_x+\ket{\downarrow_{t=0}}_x\ket{\downarrow_0}_x]\,,
}
which is just a direct product of Bell pairs, one each at every site. 
Considering a cut $l$ then is equivalent to cutting through $2l$ of these Bell pairs which immediately leads to ${\cal P}(l,t=0)=\exp[-2|l|\ln 2]$, this is shown schematically in Fig.~\ref{fig:schematic-operator-entanglement}(a).
At $t>0$, the spins at site $x$ get entangled with its neighbours within some distance, bounded by the depth of the unitary, from $x$ such that the ${\cal P}(l,t)$ decays from its initial value, or equivalently, $S_2^{AB}(t)$ grows; this is shown schematically in Fig.~\ref{fig:schematic-operator-entanglement}(b).
Finally, another trivial limiting case is that of $|l|=L/2$ in which case one subsystem is composed of all the spins at $t=0$ and the other, at $t=t$. In this case, the unitarity of $U_t$ dictates that ${\cal{P}}(\pm L/2, t) = \exp[-L \ln 2]$ at all $t$. 
A qualitative picture therefore emerges for ${\cal P}(l,t)$ on completely general grounds which is shown schematically in Fig.~\ref{fig:schematic-operator-entanglement}(c). Making this picture quantitative and showing the emergence of a logarithmic lightcone therein from eigenstate correlations in the MBL phase will constitute the subsequent sections of the paper.

\subsection{opEE and eigenstate correlations \label{sec:eigcorr}}
With the definitions of the opEE of $U_t$ at hand, we now discuss how are they encoded in the eigenstate correlations of the generator of time-translation. We will denote the time-evolution operator over one unit of time as $U_F$, such that $U_t=U_F^t$, and 
\eq{
U_F = \sum_{\alpha}e^{-i\theta_\alpha}\ket{\alpha}\bra{\alpha}\,,
}
where $\ket{\alpha}$ is an eigenstate of $U_F$ with eigenvalue $e^{-i\theta_\alpha}$.
To relate the opEE or the operator purity of $U_t$ (see Eq.~\ref{eq:S2-opEE} and Eq.~\ref{eq:pu-opEE}) to eigenstate correlations, let us define sets of basis states $\{\ket{i_{A_{0/t}}}\}$ and $\{\ket{i_{B_{0/t}}}\}$ for subsystems $A_{0/t}$ and $B_{0/t}$.
Note that a state on the physical $L$ qubits, such as an eigenstate $\ket{\alpha}$ of $U_F$ can be written in the basis at $t=0$ or equivalently at $t=t$ as 
\eq{
\ket{\alpha} = \sum_{i_{A_0}i_{B_0}}\alpha_{i_{A_0}i_{B_0}}\ket{i_{A_0}i_{B_0}}=\sum_{i_{A_t}i_{B_t}}\alpha_{i_{A_t}i_{B_t}}\ket{i_{A_t}i_{B_t}}\,.
}
With this notation, the state representation of the operator $U_t$ becomes
\eq{
|U_t) = \frac{1}{\sqrt{\nh}}&\sum_{\alpha}e^{-i\theta_\alpha t}\times\nonumber\\&\sum_{\substack{i_{A_0},i_{B_0}\\i_{A_t},i_{B_t}}}\alpha_{i_{A_t}i_{B_t}}\alpha^\ast_{i_{A_0}i_{B_0}}\ket{i_{A_t}i_{B_t}}\ket{i_{A_0}i_{B_0}}\,.
\label{eq:Ut-state-basis}
}
The form in Eq.~\ref{eq:Ut-state-basis} makes the interpretation of the operator $U_t$ as a state on $2L$ spins manifest, and one can compute entanglement across any bipartition of these $2L$ spins. In particular for the spatiotemporal bipartition parametrised by $l$, (see Fig.~\ref{fig:schematic-operator-entanglement}) the operator purity can be expressed as (see Appendix \ref{app:operator-purity} for details)
\eq{
{\cal P}(l,t)=\frac{1}{\nh^2}\sum_{\alpha\beta\gamma\lambda}e^{-it\theta_{\alpha\beta\gamma\lambda}}V_{\alpha\beta\gamma\lambda}^{(l)}
\left(V_{\alpha\beta\gamma\lambda}^{(-l)}\right)^\ast\,,
\label{eq:Plt-eigcorr}
}
where $\theta_{\alpha\beta\gamma\lambda}=\theta_\alpha-\theta_\beta-\theta_\gamma+\theta_\lambda$ and 
\eq{
V_{\alpha\beta\gamma\lambda}^{(l)} = \sum_{\substack{i_X,i_{\overline{X}},\\ j_X,j_{\overline{X}}}} \alpha_{i_Xi_{\overline{X}}}
\beta^{\ast}_{j_Xi_{\overline{X}}}
\gamma_{i_Xj_{\overline{X}}}^{\ast}
\lambda_{j_Xj_{\overline{X}}},\,,
\label{eq:V-abcd-l}
}
with $X$ the subsystem containing the leftmost $L/2-l$ sites, ${\overline{X}}$ its complement,
and $\{\ket{i_X}\}~(\{\ket{i_{\overline{X}}}\})$ denoting a set of basis state for the former (latter).
While Eq.~\ref{eq:Plt-eigcorr} pertains to the space-time structure of the opEE, it will also be useful to study it in the frequency domain
\eq{
{\tilde{\cal P}}(l,\omega) = \frac{1}{\nh^2}\sum_{\alpha\beta\gamma\lambda}\delta_{2\pi}(\omega-\theta_{\alpha\beta\gamma\lambda})V_{\alpha\beta\gamma\lambda}^{(l)}
\left(V_{\alpha\beta\gamma\lambda}^{(-l)}\right)^\ast\,,
\label{eq:Plomega-eigcorr}
}
which is simply a Fourier transform of Eq.~\ref{eq:Plt-eigcorr}. In the following, for brevity of notation, we will use
${\cal V}^{(l)}_{\alpha\beta\gamma\lambda} = V_{\alpha\beta\gamma\lambda}^{(l)}
\left(V_{\alpha\beta\gamma\lambda}^{(-l)}\right)^\ast$.
Note that Eq.~\ref{eq:Plt-eigcorr} can also be interpreted as the average over the OTOCs~\cite{hahn2023statistical}

\eq{
{\cal P}(l,t) = \frac{1}{D_XD_Y\nh^2}\sum_{O_X,O_Y}{\rm Tr}[O_X(t)O_YO_X(t)O_Y]\,,
}
where $\{O_X\}$ denotes the set of all the $D_X^2$ operators supported on subsystem $X$ comprising the sites $1$ through $L/2-l$ (with $D_X$ its Hilbert space) and similarly for $\{O_Y\}$ supported on subsystem $Y$ containing the sites $L/2+l+1$ through $L$.
The results in Eq.~\ref{eq:Plt-eigcorr} and Eq.~\ref{eq:Plomega-eigcorr} constitute a concrete relation between the spatiotemporal profile of information spreading and eigenstate correlations. In the following, we will exploit a hierarchical structure in the latter for MBL systems~\cite{pain2024entanglement} to develop a microscopic theory for the logarithmic entanglement lightcone in such systems.

\section{Disordered Floquet Ising Spin Chain }\label{sec: Num-Floquet-Ising}

\begin{figure}    \includegraphics[width=\linewidth]{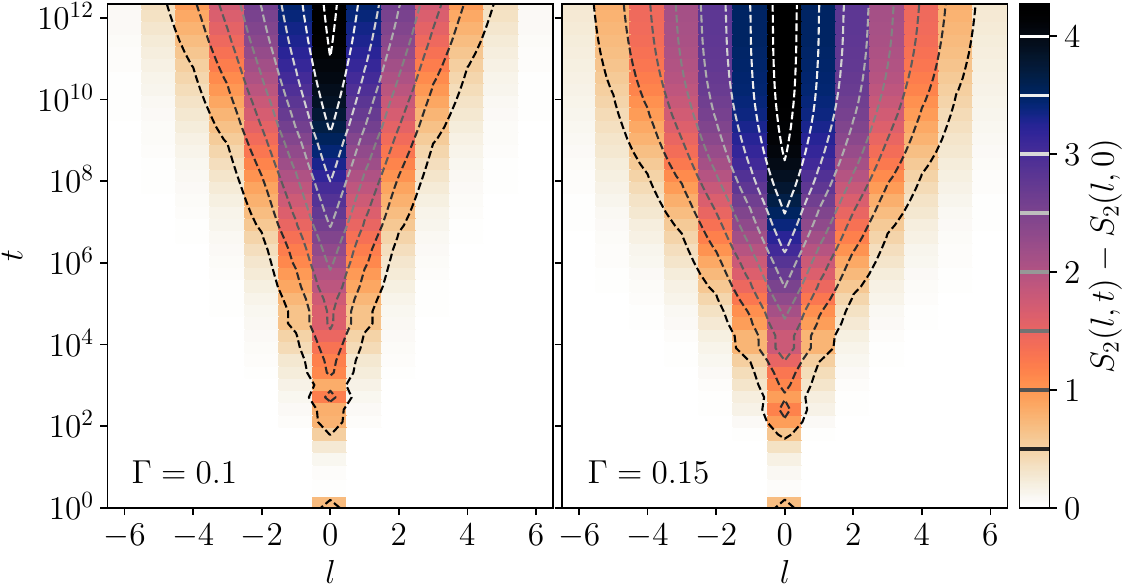}
    \caption{Spatiotemporal profile of the opEE (second R\'enyi entropy) of $U_t$, relative to its initial profile, $S_2(l,t)-S_2(l,0)$, as a heatmap in the $l$-$t$ plane shows the logarithmic lightcone. Note that the time axis is on logarithmic scales and the dotted lines denote contours of fixed  $S_2(l,t)-S_2(l,0)$ at values marked in the colourbar. The results are for the disordered Floquet Ising model in Eq.~\ref{eq:UF-TFI} with the two panels corresponding to $\Gamma=0.1$ and $0.15$, representative of the MBL regime. Note that for smaller $\Gamma$ (deeper in the MBL phase), the logarithmic lightcone is visibly slower. The data is for $L=14$.}
\label{fig:log-lightcone}
\end{figure}

To discuss our results and demonstrate the theory on a concrete footing we consider a disordered, interacting Floquet Ising spin chain, the $U_F$ for which is given by 
\eq{
U_F = \exp[-i\tau H_X] \exp[-i\tau H_Z]\,,
\label{eq:UF-TFI}
}
with 
\eq{
\begin{split}
    H_X &= g\Gamma \sum_{i=1}^L \sigma^x_i\,, \\
    H_Z &= \sum_{i=1}^L \big[\sigma^z_i \sigma^z_{i+1} + \big(h + g\sqrt{1-\Gamma^2}\epsilon_i\big)\sigma^z_i\big]\,,
\end{split}}
where $\{\sigma_i^\mu\}$ represents the Pauli matrices corresponding to the spins-$1/2$, and $\epsilon_i \sim \mathcal{N}(0,1)$ are independent Gaussian random variables with zero mean and unit variance. Following Ref.~\cite{zhang2016floquet}, we take the parameters $g = 0.9045$, $h = 0.809$, and $\tau = 0.8$. For these parameters, the model exhibits a putative many-body localised phase for $\Gamma\lessapprox 0.3$. All our numerical results will therefore be for $\Gamma = 0.1$ and $\Gamma = 0.15$, two representative values within the MBL phase and are averaged over 50-100 disorder realisations.

\begin{figure}
\includegraphics[width=\linewidth]{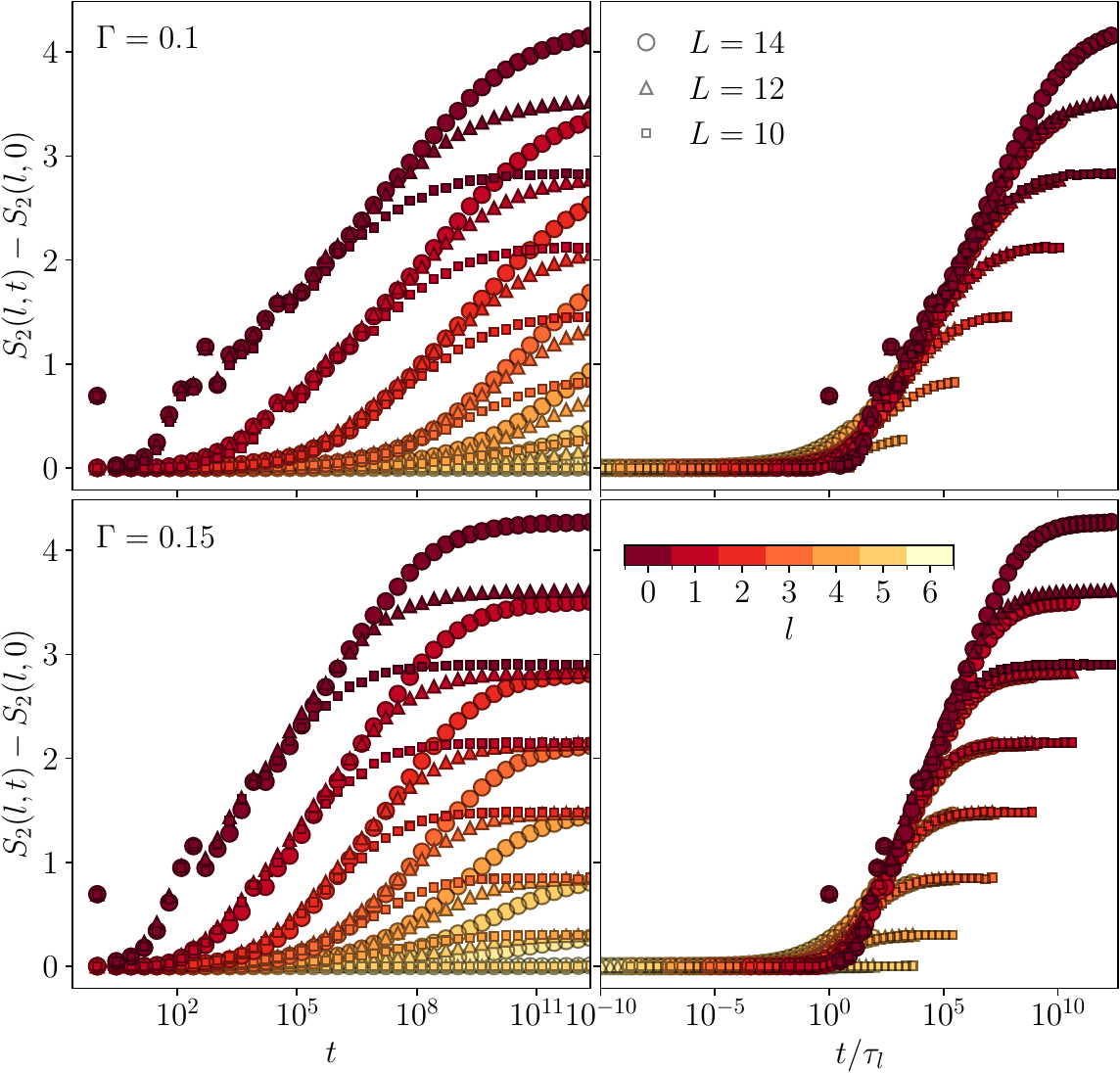}
    \caption{The opEE of $U_t$ for different spatial cuts, $l$, as a function of time $t$. The left panels shows the raw data for $S_2(l,t)-S_2(l,0)$ whereas in the right panels, the time axis has been rescaled by $\tau_l$ (see Eq.~\ref{eq:tstar-l}) which collapses the onset of the opEE for different $l$. The exponential scaling of $\tau_l$ with $l$ (see Eq.~\ref{eq:tstar-l}) implies the presence of a logarithmic lightcone. Different colours denote different values of $l$ whereas different markers denote different system sizes $L$. The top and bottom rows correspond to $\Gamma=0.1$ and $0.15$ respectively.}
    \label{fig:S_2(l,t) scaling}
\end{figure}

To establish the basic phenomenology of the spatiotemporal profile of the opEE of $U_t$, we show, in Fig.~\ref{fig:log-lightcone}, $S_2(l,t)$ relative to its $t=0$ profile as a heatmap in the $(l,t)$ plane. It is immediately apparent from the data that there exists a logarithmically spreading lightcone of $S_2(l,t)$.
To get a more quantitative picture we show, in the left panels, of Fig.~\ref{fig:S_2(l,t) scaling}, the data for $S_2(l,t)-S_2(l,0)$ as a function of $t$ for different values of $l$. The results therein make the logarithmic lightcone extremely evident as $S_2(l,t)-S_2(l,0)\approx 0$ until some characteristic timescale $\tau_l$ and picks up finite values only after that. Formally, in the limit of $L\gg 1$, the profile of $S_2(l,t)$ can be expressed as 
\eq{S_2(l,t)=\begin{cases}
    S_2(l,0)\,; & t\ll \tau_l \\
    S_2(l,0) + c_1\ln t\,; & t\gg \tau_l
\end{cases}\,,\label{eq: S_2(l,t)-scaling}}  
where $S_2(l,0) = 2|l|\ln 2$.
Scaling $t$ by $\tau_l$ collapses the data for the onset of $S_2(l,t)-S_2(l,0)$ for different values of $l$ as shown in the right panels of Fig.~\ref{fig:S_2(l,t) scaling}. 
More importantly, the scaling collapse is achieved for 
\eq{
\tau_l = \tau_0\exp[l/\zeta]\,,
\label{eq:tstar-l}
}
where $\zeta$ is a lengthscale which naturally depends on the microscopic parameters of the model. We find that $\zeta$ decays with moving deeper into the MBL phase, manifested in the logarithmic lightcone being visibly slower for smaller $\Gamma$ in Fig.~\ref{fig:log-lightcone}. Specifically, we find that $\zeta=0.19$ for $\Gamma=0.1$, and $\zeta=0.25$ for $\Gamma=0.15$.

However, the important upshot of Eq.~\ref{eq: S_2(l,t)-scaling} and Eq.~\ref{eq:tstar-l} is that they immediately imply the existence of a logarithmic lightcone in $S_2(l,t)$. 
Much of the remainder of this paper will be about developing a microscopic theory for the scaling of the onset time 
$\tau_l$ and the subsequent logarithmic entanglement lightcone. 
Note that the `universal' form in Eq.~\ref{eq: S_2(l,t)-scaling} will be cutoff by the saturation of $S_2(l,t)-S_2(l,0)$ for finite systems. In particular, it is expected, due to dephasing, that the infinite time opEE will be given by $S_2(l,t\to\infty)\propto (L/2+|l|)\ln 2$ which in turn suggests a saturation timescale of $t_{\rm sat}(l) \propto \tau_l\exp[(L/2+|l|)\ln 2]$.

\section{Logarithmic lightcone from eigenstate correlations \label{sec:log-lightcone-eigcorr}}

In this section we develop a theoretical understanding of the spatiotemporal profile of the opEE through the lens of eigenstate correlations using the explicit relation in Eq.~\ref{eq:Plomega-eigcorr}.
In particular, we show that from the hierarchy of energyscales contained in the eigenstate correlations~\cite{pain2024entanglement}, the logarithmic lightcone emerges naturally.

\subsection{Special quartets of eigenstates \label{sec:special-quartets}}

\begin{figure}
    \centering
    \includegraphics[width=\linewidth]{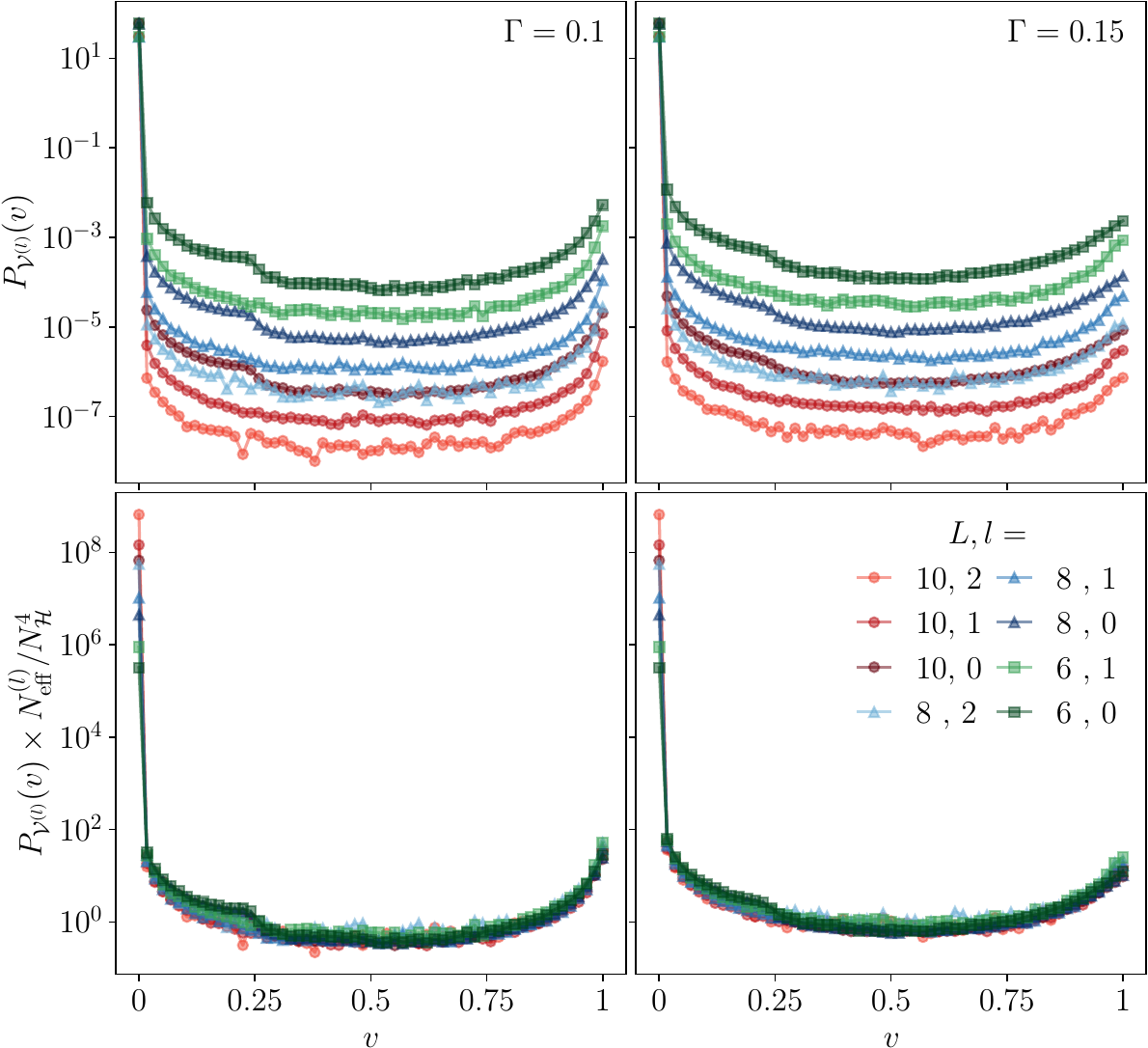}
    \caption{The distribution $P_{{\cal V}^{(l)}}(v)$ of $\Gabcd$ defined in Eq.~\ref{eq:Gabcd-dist}. The top panels show the raw distributions for different $l$ (different colours) and for different system sizes $L$ with darker shades denoting larger $L$. The bottom panels show the same distributions but rescaled by the fraction $N_{\rm eff}^{(l)}/\nh^4 \approx \nh^{2}2^{2l}$. This rescaling collapses the distribution for different $l$ and $L$ in the regime where $v\sim O(1)$.}
    \label{fig:Gabcd-dist}
\end{figure}

\begin{figure}
    \centering
    \includegraphics[width=\linewidth]{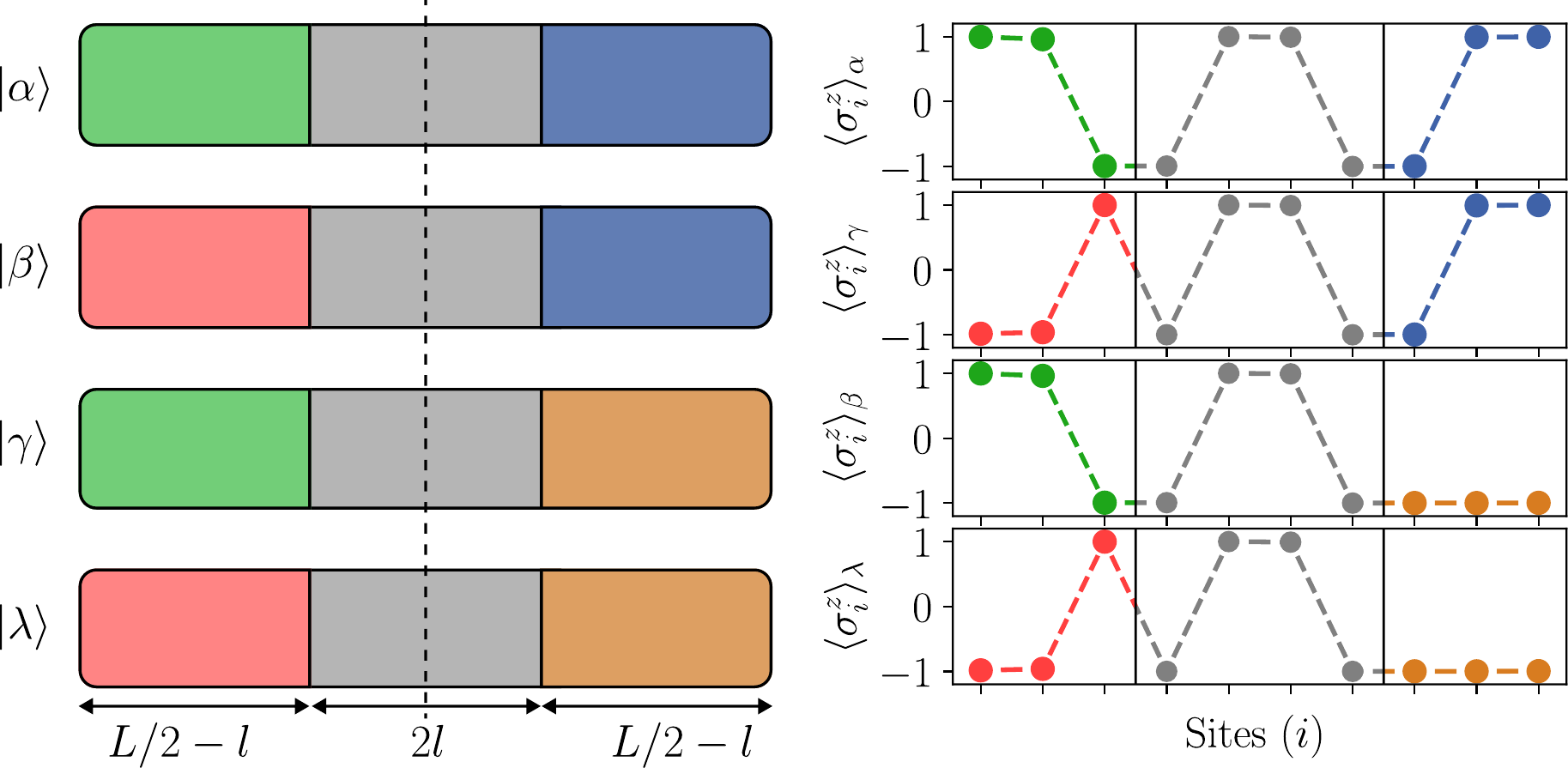}
    \caption{Real-space structure of a representative quartet of eigenstates when $\Gabcd \sim O(1)$. (Left) Schematic representation of the real-space profile of these four eigenstates. All four eigenstates share the same spin profile at the centre of the chain on sites $(L/2-l,L/2+l]$. However, in the rest of the chain, $\{\alpha, \beta\}$ and $\{\gamma, \lambda\}$ exhibit identical profiles in the subsystem $[1,L/2-l]$, while $\{\alpha, \gamma\}$ and $\{\beta, \lambda\}$ have identical profiles in the subsystem $(L/2+l,L]$. Similar (different) spin profiles are indicated by the same (different) colour. (Right) Quantitative demonstration of the structure using the real-space profile of $\sigma^z_i$ in the four eigenstates for a system with $L=10$ and $l = 2$.}
    \label{fig: Gabcd-schematic}
\end{figure}

Given Eq.~\ref{eq:Plomega-eigcorr}, it is clear that $\Gabcd$ is a central quantity of interest. We therefore start with the statistics of $\Gabcd$ as encoded in its probability distribution,

\eq{
P_{\mathcal{V}^{(l)}}(v)=\frac{1}{N_\mathcal{H}^4}\sum_{\al\be\ga\la}\delta(\Gabcd-v)\,.
\label{eq:Gabcd-dist}
}
Numerical results for $P_{\mathcal{V}^{(l)}}(v)$ are shown in Fig.~\ref{fig:Gabcd-dist} for two parameter values, both in the MBL phase. 
The key point to note that while an overwhelming majority of the quartets have $\Gabcd\approx 0$, an extremely small fraction of the quartets have $\Gabcd\sim O(1)$. The fraction of the latter also appears to decrease with increasing system size and $l$. In the rest of what follows, we will refer to these quartets as the special dominant quartets.
As a matter of notation we will denote the set of these special dominant quartets of eigenstates dominating the opEE of $U_t$ for the cut as $\square_l$,
\eq{
\Gabcd\sim O(1)~\forall (\alpha,\beta,\gamma,\delta)\in \square_l\,.
}
To gain an understanding of the structure of the eigenstates which constitute these special quartets, it is useful to consider the form of $\Vabcdl{l}$ in Eq.~\ref{eq:V-abcd-l}, particularly in the $\sigma^z$-product state basis. The latter is motivated by the fact that in the limit of strong disorder, each MBL eigenstate, notwithstanding its multifractal nature on the Hilbert-space, can be associated with a well-defined localisation centre on the Hilbert-space which is nothing but a $\sigma^z$-product state. The form in Eq.~\ref{eq:V-abcd-l} then suggests that for $\Vabcdl{\pm l}$ to have an $O(1)$ magnitude, 
\begin{enumerate}
    \item[(i)] the localisation centres of the pair $\ket{\al}$ and $\ket{\ga}$, and the pair $\ket{\be}$ and $\ket{\la}$ should be similar in the subsystem comprising sites 1 through $L/2\mp l$, and 
    \item[(ii)] the localisation centres of the pair $\ket{\al}$ and $\ket{\be}$, and the pair $\ket{\ga}$ and $\ket{\la}$ should be similar in the subsystem comprising sites $L/2\mp l$ through $L$.
\end{enumerate}
For $\Gabcd$ to be of magnitude $O(1)$ we require both $\Vabcdl{l}$ and $\Vabcdl{-l}$ to have an $O(1)$ magnitude simultaneously. The above two conditions which ensure that can be formally written as
\eq{
\begin{split}
    &\braket{\z_i}_\alpha = \braket{\z_i}_\gamma, \quad \braket{\z_i}_\beta = \braket{\z_i}_\lambda, \, \forall i \in [1, L/2-l] \\
    &\braket{\z_i}_\alpha = \braket{\z_i}_\beta, \quad \braket{\z_i}_\gamma = \braket{\z_i}_\lambda, \, \forall i \in (L/2+l, L] \\
    &\braket{\z_i}_\alpha = \braket{\z_i}_\beta = \braket{\z_i}_\gamma = \braket{\z_i}_\lambda, \, \forall i \in (L/2-l, L/2+l]
\end{split}\,.
\label{eq:sz-cond}
}
This is shown schematically in the left panel of Fig.~\ref{fig: Gabcd-schematic} where the three sections of the chain correspond to sites $[1,L/2-l]$, $(L/2-l,L/2+l]$, and $(L/2+l,L]$, and sections of the chain having the same colour denote the same profile of $\sigma^z_i$ for the localisation centres. In the right panel of Fig.~\ref{fig: Gabcd-schematic}, the $\braket{\sigma^z_i}$ profiles for the four eigenstates in a randomly chosen quartet with $\Gabcd\sim O(1)$ is shown which confirms the conditions laid out in Eq.~\ref{eq:sz-cond}.

The aforementioned anatomy of the eigenstates forming the special quartets lends itself to a computation of their multiplicity as well. Consider the the left panel in Fig.~\ref{fig: Gabcd-schematic}. For a particular $\ket{\alpha}$ of which there are $2^L$ choices, the configuration of $\ket{\beta}$ in the right and middle sections of the chain are also fixed but there are $2^{L/2-l}$ choices for the left section. Similarly, configuration of $\ket{\gamma}$ in the left and middle sections are fixed by there are $2^{L/2-l}$ choices for the right section. Once the $\ket{\alpha}$, $\ket{\beta}$ and $\ket{\gamma}$ are fixed, it automatically fixes the state $\ket{\lambda}$. The total number of such special quartets, contributing $\Gabcd\sim O(1)$ to the opEE for the cut at $l$ can therefore be estimated as
\eq{
N_\text{eff}^{(l)} \sim 2^L \times 2^{L/2-l} \times 2^{L/2-l} =\nh^2 2^{-2l}\,,
}
or equivalently, the fraction of all quartets with $\Gabcd\sim O(1)$ is $N_\text{eff}^{(l)}/\nh^4\sim \nh^{-2} 2^{-2l}$.
Quantitative evidence for this counting is presented in the lower panels of Fig.~\ref{fig:Gabcd-dist} where scaling the distribution, $P_{{\cal V}^{(l)}}(v)$ by $N_{\rm eff}^{(l)}$ collapses the part of the distribution with $v\sim O(1)$ for several $L$ and $l$.

\subsection{Spectral correlations within the special dominant quartets \label{sec:spect-corr}}

\begin{figure}
    \centering
    \includegraphics[width=\linewidth]{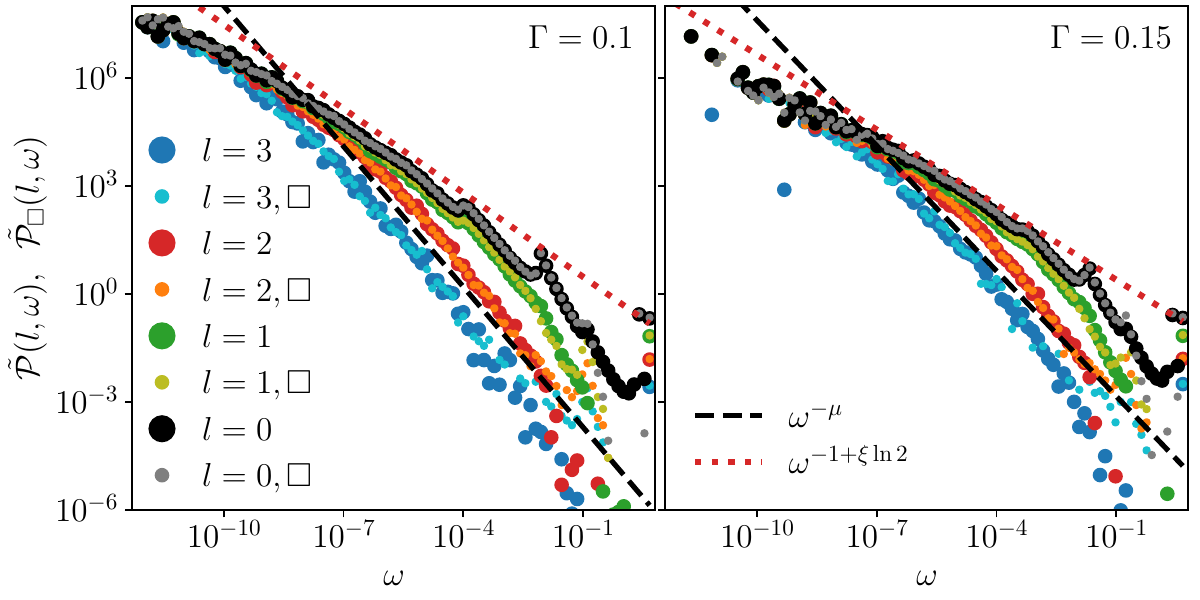}
    \caption{Comparison between ${\tilde{\cal P}}(l,\omega)$ (defined in Eq.~\ref{eq:Plomega-eigcorr}) and $\pos$ (defined in Eq.~\ref{eq:Pbox-omega}) shows that they are in excellent agreement with each other. This provides numerical evidence for the validity of the approximation made in Eq.~\ref{eq:Pbox-omega} which allows the opertor purity of $U_t$ to be expressed purely in terms of spectral correlations within the special dominant quartets. The dashed and dotted lines denote the two-power law behaviours of ${\tilde{\cal P}}(l,\omega)$ with $\omega$ mentioned in Eq.~\ref{eq:Plomega-final-result}.}
    \label{fig:Pomega-Pboxomega-compare}
\end{figure}

The identification of the special dominant quartets, in particular their anatomy and multiplicity, suggests that the opEE of $U_t$ can be understood entirely from the spectral correlation within them. To see this note the identity
\eq{
\sum_{\alpha\beta\gamma\lambda}\Gabcd = \nh^2 2^{-2l} \sim N_{\rm eff}^{(l)}\,,
}
which suggests that the operator purity in Eq.~\ref{eq:Plomega-eigcorr} is indeed dominated entirely by the special quartets 
\eq{
{\tilde{\cal P}}(l,\omega) \approx \frac{1}{\nh^2}\sum_{\alpha\beta\gamma\lambda\in \square_l} \delta_{2\pi}(\tabcd-\omega)\Gabcd\,.
}
In addition, given that $\Gabcd\sim O(1)$ for all such quartets the operator purity is extremely well approximated by the spectral correlations within the special quartets~\cite{pain2024entanglement}
\eq{
{\tilde{\cal P}}(l,\omega) \approx \pos \equiv  \frac{1}{\nh^2} \sum_{\alpha \beta \gamma \lambda \in \square_l} \delta(\tabcd - \omega)\,.
\label{eq:Pbox-omega}
}
Numerical evidence for the same is shown in Fig.~\ref{fig:Pomega-Pboxomega-compare}.
The key physical import of the above is that the operator purity is dominated by the eigenstate correlations within a vanishing fraction of special dominant quartets and their anatomy is such that the operator purity is governed simply by the spectral correlations within these quartets.

\begin{figure}[!t]
\includegraphics[width=\linewidth]{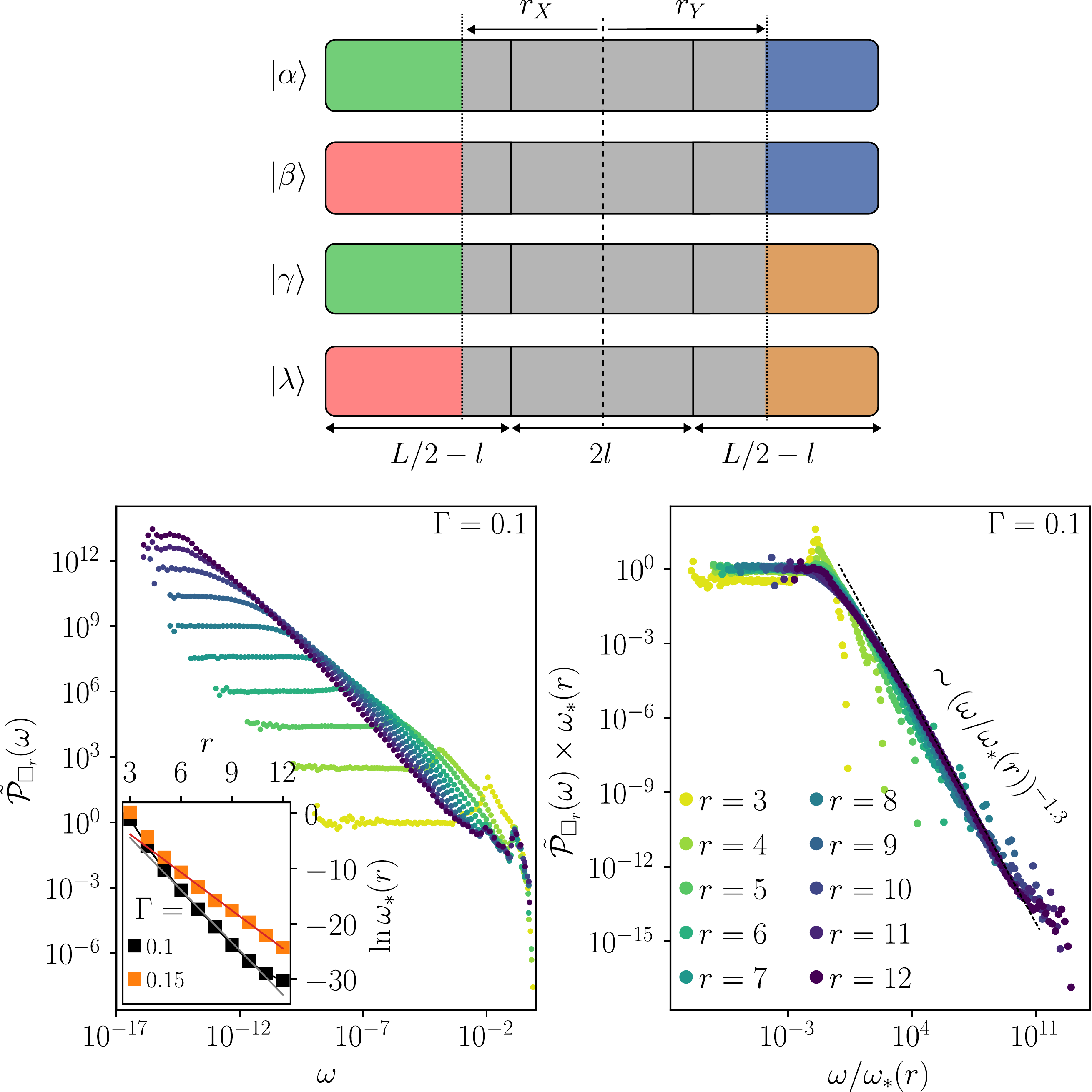}
    \caption{Top: Schematic representation of the quartets of eigenstates which constitute the set $\square_{r_Xr_Y}$. As in Fig.~\ref{fig: Gabcd-schematic}, similar (different) colours indicate the localisation centres of the corresponding subsystems have similar (different) $\sigma^z$-profiles. Bottom: Spectral correlations within the quartets in $\square_{r_Xr_Y}$ and their scaling collapse showing evidence for the scaling form for $\posrArB$ given in Eq.~\ref{eq:Pomega-scaling}. This is the same data that we presented in Ref.~\cite{pain2024entanglement}, and to avoid repetition we only show $\Gamma=0.1$.}
    \label{fig:rArB-PrArB}
\end{figure}

Note that the anatomy of the eigenstates in $\square_l$ shown in Fig.~\ref{fig: Gabcd-schematic} implies that they form a subset of the special dominant quartets which govern the entanglement dynamics for $l=0$, with the latter case studied in detail in Ref.~\cite{pain2024entanglement}. This naturally allows us to exploit the results obtained therein for the spectral correlations within the dominant quartets. 
{In particular, consider the following characterisation of the special quartets. The structure of the special quartets mandates that the spin-configurations are identical for the sites $i\in(L/2-l,L/2+l]$. However, the quartets can be further classified based on where do the states differ pairwise, outside this central region. We will denote by $L/2-r_X$, the location of the rightmost spin which is necessarily different between in the pair $\ket{\alpha}$ and $\ket{\beta}$, and therefore automatically in the pair $\ket{\gamma}$ and $\ket{\lambda}$. Similarly, we will denote by $L/2+r_Y$ the location of the leftmost spin  which is necessarily different between in the pair $\ket{\alpha}$ and $\ket{\gamma}$, and therefore automatically in the pair $\ket{\beta}$ and $\ket{\lambda}$. We will label the set of all such quartets as $\square_{r_Xr_Y}$. Their defining features (see Fig.~\ref{fig:rArB-PrArB}(top) for a visual representation) can be summarised as 
\begin{enumerate} 
    \item[(i)] the localisation centres of the pair $\ket{\alpha}$ and $\ket{\beta}$, and of the pair $\ket{\gamma}$ and $\ket{\lambda}$ are similar for subsystem comprising sites $L/2-r_X$ through $L$, and 
    \item[(ii)] the localisation centres of the pair $\ket{\alpha}$ and $\ket{\gamma}$, and of the pair $\ket{\beta}$ and $\ket{\lambda}$ are identical for subsystem comprising sites $1$ through $L/2+r_Y$.
    \item[(iii)] the spins at the interface, namely at $L/2-r_X$ and $L/2+r_Y$, are necessarily different between the pairs $(\alpha,\beta)$ and $(\gamma,\lambda)$ for the former, and the pairs $(\alpha,\gamma)$ and $(\beta,\lambda)$ for the latter.
\end{enumerate}}
The operator purity, from Eq.~\ref{eq:Pbox-omega} is then given by~\footnote{{Note a subtle detail in the notation here. Our notation implies that the in quartets $\square_{r_Xr_Y}$, the spins at sites $L/2-r_X$ and $L/2+r_Y$ are necessarily different between the respective pairs. However, since the summations in Eq.~\ref{eq:pomega-sum} run only up to $L/2$, they exclude the possibility of the eigenstates being pairwise the same, or all four of of them being same. This in turn means that Eq.~\ref{eq:pomega-sum} is valid strictly for $\omega\neq 0$. Notwithstanding this issue, we continue to use Eq.~\ref{eq:pomega-sum} as we are only interested in the dynamical regime which does correspond to $\omega\neq 0$.}}
\eq{
\pos = \frac{1}{\nh^2}\sum_{r_X=l+1}^{L/2}\sum_{r_Y=l+1}^{L/2}N_{r_Xr_Y}\posrArB\,,
\label{eq:pomega-sum}
}
where $\posrArB$ is the normalised spectral correlations within the eigenstates in $\square_{r_Xr_Y}$,
\eq{
\posrArB=\frac{1}{N_{r_Xr_Y}}\sum_{\substack{\al\be\ga\la\in\square_{r_Xr_Y}}}\delta(\tabcd-\omega)\,.
}
with 
\eq{
{N_{r_Xr_Y} = 2^{2L -r_X - r_Y}}\,,
\label{eq:Nrxry}
}
the number of quartets in $\square_{r_Xr_Y}$. This expression for $N_{r_Xr_Y}$ can be understood as follows from Fig.~\ref{fig:rArB-PrArB}(top). For a given $\ket{\alpha}$  for which there are $\nh$ choices, the localisation centres of $\ket{\beta}$ and $\ket{\gamma}$ can be different from that of $\ket{\alpha}$ only over $L-r_X$ and $L-r_Y$ sites respectively, and fixing $\ket{\alpha}$, $\ket{\beta}$ and $\ket{\gamma}$ fixes $\ket{\lambda}$. The multiplicity of quartets $\square_{r_Xr_Y}$ is therefore $\nh\times 2^{L-r_X}\times 2^{L-r_Y}$ which the result in Eq.~\ref{eq:Nrxry}.

It was shown in Ref.~\cite{pain2024entanglement} that $\posrArB$ admits a scaling form 
\eq{
\posrArB = \omega^{-1}_\ast(r)\mathcal{F}\bigg(\frac{\omega}{\omega_\ast(r)}\bigg)\,;\quad r=r_X+r_Y\,,
\label{eq:Pomega-scaling}
}
with the asymptotics of the scaling function given by
\eq{
    \mathcal{F}(x) = \begin{cases}
         1\,; & x \to 0 \\
            x^{-\mu}\,; & x \gg 1
    \end{cases}\,,
    \label{eq:asymptotic}
}  
where $\mu>1$ necessarily.
It was also shown in Ref.~\cite{pain2024entanglement} that the characteristic energyscale $\omega_\ast(r)$ scaled with $r$
\eq{\omega_\ast(r) = c \exp(-r/\xi)\,,\label{eq:omega-ast}}
where $\xi$ is an emergent lengthscale which naturally depends on the microscopic parameters of the model but satisfies the inequality $\xi \ln 2<1$ throughout the MBL phase. For the sake of completeness, we reproduce the data for $\posrArB$ from Ref.~\cite{pain2024entanglement} in Fig.~\ref{fig:rArB-PrArB}(bottom), which provides credence to the scaling form in Eq.~\ref{eq:Pomega-scaling}.
It then remains only to evaluate the sum in Eq.~\ref{eq:pomega-sum} using the results in Eq.~\ref{eq:Pomega-scaling} through Eq.~\ref{eq:omega-ast} to derive a result for ${\tilde{\cal P}}(l,\omega)$, which we do in the next subsection.

\subsection{Spatiotemporal profile of the opEE \label{sec:log-lc-deriv}}
With the expressions for $\posrArB$ in Eq.~\ref{eq:Pomega-scaling} and $N_{r_Xr_Y}$ in Eq.~\ref{eq:Nrxry} at hand, we can now evaluate ${\tilde{\cal{P}}}(l,\omega)$ by explicitly evaluating the sum in Eq.~\ref{eq:pomega-sum}.
Inspection of the scaling form of $\posrArB$ in Eq.~\ref{eq:Pomega-scaling} and the form of $\omega_\ast(r)$ in Eq.~\ref{eq:omega-ast} immediately suggests the presence of a characteristic energyscale for a given $l$,
\eq{
\Omega_l \equiv \omega_\ast(2l)= c \exp({-2l/\xi})\,.
\label{eq:Omega-l}
}

For $\omega\gg \Omega_l$, all the summands in Eq.~\ref{eq:pomega-sum} correspond to $\square_{r_Xr_Y}$ for which $\omega$ is greater than the corresponding $\omega_\ast(r_X+r_Y)$ and the scaling function ${\cal F}(x)$ in Eq.~\ref{eq:Pomega-scaling} follows the second line in Eq.~\ref{eq:asymptotic}. We therefore have
\eq{
\pos &\overset{\omega\gg\Omega_l}{=} \omega^{-\mu}c^{-1+\mu}\smashoperator{\sum_{r_X,r_Y=l+1}^{L/2}}e^{-(r_X+r_Y)[\frac{\mu-1}{\xi}+\ln 2]}\nonumber\\
&\overset{L\gg 1}{\approx}\left(\frac{\omega}{\Omega_l}\right)^{-\mu} \frac{c^{-\xi \ln 2}\Omega_l^{-1+\xi\ln 2}}{\left(e^{\frac{\mu-1}{\xi}+\ln 2}-1\right)^2}
\label{eq:pomega>Omegal}
}
where we have explicitly used the form of $\omega_\ast(r)$ from Eq.~\ref{eq:omega-ast} and $N_{r_Xr_Y}$ from Eq.~\ref{eq:Nrxry}. The result can be understood as, since $\mu>1$ necessarily~\cite{pain2024entanglement} the summands decay exponentially with $r_X,r_Y$ and therefore the sum is dominated by values of $r_X,r_Y$ close to $l+1$.

For the opposite case of $\omega\ll\Omega_l$, in the summands in Eq.~\ref{eq:pomega-sum}, the scaling function in $\posrArB$ follows the first line of Eq.~\ref{eq:Pomega-scaling} if $r=r_X+r_Y > r_\ast(\omega)$ with $r_\ast(\omega)=\xi\ln|\omega/c|$ or the second line if $r < r_\ast(\omega)$. The sum can therefore be split as 
\begin{widetext}
\eq{
\pos &\overset{\omega\ll\Omega_l}{=}~c^{-1}\smashoperator{\sum_{\substack{r_X,r_Y:\\r_X,r_Y\ge l+1\\r_X+r_Y\leq r_\ast(\omega)}}}e^{-(r_X+r_Y)[-\frac{1}{\xi}+\ln 2]}~+~\omega^{-\mu}c^{-1+\mu}\smashoperator{\sum_{\substack{r_X,r_Y:\\r_X+r_Y>r_\ast(\omega)\\r_X,r_Y\leq L/2}}}e^{-(r_X+r_Y)[\frac{\mu-1}{\xi}+\ln 2]}\nonumber\\
&\overset{L\gg 1}{\approx}C_0+ C_1\omega^{-1+\xi\ln 2} + C_2\omega^{-1+\xi\ln 2}\ln(\Omega_l/\omega)\,,
}
\end{widetext}
where the logarithmic correction in the last term is due to entropic considerations of pairs of $r_X$ and $r_Y$ which sum up to the same $r$, and the expressions for $C_0$, $C_1$, and $C_2$ which depend only on $\mu$ and $\xi$ are given by
\eq{
\begin{split}
C_0 &= \frac{c^{-1}e^{-2l(-\frac{1}{\xi}+\ln 2)}}{{\left(e^{-\frac{1}{\xi}+\ln 2}-1\right)^2}}\,,\\
C_1 &= \frac{c^{-\xi\ln 2}}{(e^{\frac{\mu-1}{\xi}+\ln2}-1)^2}-\frac{c^{-\xi\ln 2}}{(e^{\frac{-1}{\xi}+\ln2}-1)^2}\,,\\
C_2 &=\frac{c^{-\xi\ln 2}}{(1-e^{-\frac{1}{\xi}+\ln 2})}-\frac{c^{-\xi\ln 2}}{(1-e^{\frac{\mu-1}{\xi}+\ln 2})}\,.
\end{split}
}
In this case, $\xi\ln2<1$ and $\mu>1$ implies that in the first sum, the summands grow exponentially with $r_X,r_Y$ whereas those in the second sum decay exponentially with $r_X,r_Y$. Therefore both the sums are dominated by $r_X+r_Y\approx r_\ast(\omega)$ leading to the common $\omega^{-1+\xi\ln 2}$ behaviour. To summarise, the outcome of the above computation is the result 
\eq{
\tilde{\cal P}(l,\omega) \sim
    \begin{cases}
    \omega^{-1+\xi\ln 2}\ln(\Omega_l/\omega)\,;& \omega\ll\Omega_l\\
    (\omega/\Omega_l)^{-\mu}\,;& \omega\gg\Omega_l
    \end{cases}\,,
\label{eq:Plomega-final-result}
}
where $\xi\ln2<1$ and $\mu>1$, with $\xi$ decreasing and $\mu$ increasing on moving deeper into the MBL phase. In Ref.~\cite{pain2024entanglement}, it was found that $\xi=0.32$ and $\mu=1.30$ for $\Gamma=0.1$, and $\xi=0.44$ and $\mu=1.16$ for $\Gamma=0.15$. 
Note that the numerical results shown in Fig.~\ref{fig:Pomega-Pboxomega-compare} is entirely consistent with the form in Eq.~\ref{eq:Plomega-final-result} -- there are two power-law regimes in $\tilde{\cal P}(l,\omega)$ as a function of $\omega$ and the crossover frequency, $\Omega_l$, at which the slow power-law decay of $\sim\omega^{-1+\xi\ln 2}$ gives way to the relatively faster decay of $\sim\omega^{-\mu}$, decreases with $l$.

The result in Eq.~\ref{eq:Plomega-final-result} can be straightforwardly Fourier transformed to obtain the spatiotemporal profile of the operator purity which turns out to be
\eq{
{\cal P}(l,t) \sim {\cal P}(l,0)\times 
    \begin{cases}
    1- b(\mu,\xi) t^{\mu-1},;& t\ll\Omega_l^{-1}\\
    t^{-\xi\ln 2}\ln t\,;& t\gg\Omega_l^{-1}
    \end{cases}\,.
\label{eq:Plt-final-result}
}
The above expression clearly shows the presence of a timescale $\Omega_l^{-1} = c^{-1}e^{2l/\xi}$ across which the temporal growth of opEE of $U_t$ for the spatial cut $l$ changes qualitatively. For $t\ll \Omega_l^{-1}$, the purity decays very slowly or equivalently, the opEE grows very slowly and is approximately equal to its $t=0$ value. In the other limit of $t\gg \Omega_l^{-1}$ the second line in Eq.~\ref{eq:Plt-final-result} implies that the opEE grows logarithmically in time (modulo a further $\ln\ln t$ correction). This makes the presence of a logarithmic lightcone of entanglement manifest with the front reaching a distance $l$ at a time $t\sim \Omega_l^{-1}$ which is exponentially large in $l$. This emergent timescale can be readily identified with the timescale $\tau_l$ defined in Eq.~\ref{eq:tstar-l}. This equivalence $\tau_l \sim \Omega_l^{-1}$ implies that $\zeta \approx \xi/2$.
Comparing the values of $\zeta$ obtained from the real-time dynamics (see Fig.~\ref{fig:S_2(l,t) scaling}) and $\xi/2$ obtained from the spectral correlations within the special dominant quartets 
shows a good agreement between the two, thereby corroborating the theoretical picture.
This concludes our theoretical picture for the emergence of the logarithmic entanglement lightcone in MBL systems from the hierarchy of energyscales contained in eigenstate correlations.

\section{Conclusions \label{sec:conclusions}}
Let us summarise the paper briefly and close with some concluding remarks. In this work, we developed a microscopic theory for the logarithmic entanglement lightcone in MBL systems. As a diagnostic we used the opEE of the time-evolution operator, $U_t$, across different spatiotemporal cuts, which gave access to both spatial and temporal structure of the entanglement spreading. The theory was developed through the lens of eigenstate correlations by mapping the opEE to correlations involving four eigenstates (of $U_t$) and their eigenvalues. Interestingly enough but probably {\it a posteriori} unsurprising, the same eigenstate correlations also encode the OTOCs averaged over all operators in two distinct parts of the system. The key ingredient in the theory was the understanding that only a vanishing fraction of all the quartets eigenstates contribute to the opEE and the identification of their anatomy which of course depends on the spatiotemporal cut. In fact, the spectral correlations within these special quartets of eigenstates turn out to be sufficient to understand the opEE of $U_t$. These spectral correlations exhibit a hierarchy of energyscales and lengthscales~\cite{pain2024entanglement} from which, a characteristic timescale, $\tau_l$, emerges naturally which depends on lengthscale $l$ at which the opEE is probed -- the opEE stays close to its $t=0$ for times earlier than this characteristic timescale and starts decaying only after it. The exponential scaling $\tau_l\sim e^{l/\zeta}$ then makes the logarithmic lightcone of entanglement spreading manifest. The results presented in this paper in conjunction with our previous work~\cite{pain2024entanglement} present a complete microscopic theory for the (i) logarithmic in time growth entanglement and (ii) logarithmic in time spreading of the entanglement light cone in space in MBL systems. Interestingly, every stage in the theory was rooted in microscopic correlations and therefore the mechanism of this anomalous, ultraslow entanglement spreading in MBL systems could be understood without alluding any phenomenological constructs such as $\ell$-bits. 

With the theory and mechanism for entanglement dynamics in the MBL phase much better understood, it is natural to ask how it breaks down or needs adaptation as the MBL transition is approached. The state-of-the-art on understanding the MBL transition relies on a picture of proliferation of resonances and avalanches~\cite{morningstar2020manybody,morningstar2021avalanches,crowley2022constructive,Ha2023manybodyresonances}. However, these typically consider resonances, even if many-body, between pairs of eigenstates. On the other hand, the entanglement dynamics is controlled by correlations between quartets of eigenstates which fall out of the scope of the pairwise resonance picture. Therefore the fate of these higher-point eigenstate correlations will shed complementary and possibly much richer insights into the nature of the the MBL transition from the perspective of entanglement dynamics.

\begin{acknowledgments}
The authors thank S. Banerjee, S. Das Sarma, and A. Kundu for useful discussions. This work was supported by the Department of Atomic Energy, Government of India under Project No. RTI4001, by SERB-DST, Government of India under Grant No. SRG/2023/000858 and by a Max Planck Partner Group grant between ICTS-TIFR, Bengaluru and MPIPKS, Dresden.
\end{acknowledgments}

\appendix
\begin{widetext}
\section{Derivation of the relation between operator purity and eigenstate correlations \label{app:operator-purity}}
In this appendix we provide the details of the derivation of the expression for operator purity in Eq.~\ref{eq:Plt-eigcorr}. From the state representation of $U_t$ in Eq.~\ref{eq:Ut-state-basis}, the corresponding density matrix can be written as 

\eq{
\rho(U_t) = &\frac{1}{\nh}\sum_{\alpha\beta}e^{-i(\theta_\alpha-\theta_\beta)t}\sum_{\substack{i_{A_0},i_{B_0}\\i_{A_t},i_{B_t}}}\sum_{\substack{j_{A_0},j_{B_0}\\j_{A_t},j_{B_t}}}[ \alpha_{i_{A_t}i_{B_t}}\alpha^\ast_{i_{A_0}i_{B_0}}\beta^\ast_{j_{A_t}j_{B_t}}\beta_{j_{A_0}j_{B_0}}\ket{i_{A_t}i_{B_t}}\ket{i_{A_0}i_{B_0}}\bra{j_{A_t}j_{B_t}}\bra{j_{A_0}j_{B_0}}]\,,
}
such that tracing out the sites in $B\equiv B_0\cup B_t$ yields the reduced density matrix for $A\equiv A_0\cup A_t$ as
\eq{
\rho_A(U_t)=\frac{1}{\nh}\sum_{\alpha\beta}e^{-i(\theta_\alpha-\theta_\beta)t}\sum_{\substack{i_{A_0},j_{A_0},\\i_{A_t},j_{A_t}}}
\ket{i_{A_t}i_{A_0}}\bra{j_{A_t}j_{A_0}}\sum_{i_{B_0},i_{B_t}}\alpha_{i_{A_t}i_{B_t}}\alpha^\ast_{i_{A_0}i_{B_0}}\beta^\ast_{j_{A_t}i_{B_t}}\beta_{j_{A_0}i_{B_0}}\,.
}
From the above equation, the operator purity for the subsystem $A$ can be expressed as 
\eq{
{\rm Tr}_A[\rho_A^2(U_t)] = \frac{1}{\nh^2}\sum_{\alpha\beta\gamma\lambda}e^{-i\tabcd t}\underbrace{\sum_{\substack{i_{A_t},i_{B_t},\\j_{A_t},j_{B_t}}}\alpha_{i_{A_t}i_{B_t}}\beta^\ast_{j_{A_t}i_{B_t}}\gamma^\ast_{i_{A_t}j_{B_t}}\lambda_{j_{A_t}j_{B_t}}}_{V_{\al\be\ga\la}^{(l)}}\underbrace{\sum_{\substack{i_{A_0},i_{B_0},\\j_{A_0},j_{B_0}}}(\alpha_{i_{A_0}i_{B_0}}\beta^\ast_{j_{A_0}i_{B_0}}\gamma^\ast_{i_{A_0}j_{B_0}}\lambda_{j_{A_0}j_{B_0}})^\ast}_{(V_{\al\be\ga\la}^{(-l)})^\ast}\,.
}

For the cut parametrised by $l$ (see Fig.~\ref{fig:schematic-operator-entanglement}, $A_0\equiv[1,L/2+l]$ and $B_0$ is its complement ${\overline{A_0}}$, and $A_t\equiv[1,L/2-l]$ and $B_t$ is its complement ${\overline{A_t}}$. Identifying the structure of $V_{\alpha\beta\gamma\lambda}^{(l)}$ from Eq.~\ref{eq:V-abcd-l}, the above equation can be identified as
\eq{
{\rm Tr}_A[\rho_A^2(U_t)] = \frac{1}{\nh^2}\sum_{\alpha\beta\gamma\lambda}e^{-i\tabcd t}V_{\alpha\beta\gamma\lambda}^{(l)}(V_{\alpha\beta\gamma\lambda}^{(-l)})^\ast\,,
}
which is exactly the expression in Eq.~\ref{eq:Plt-eigcorr}.
\end{widetext}

\bibliography{refs}

\end{document}